\documentstyle[aasms4,amstex,amsfonts,epsfig,rotating,float]{article}

\def\kms{{\rm\,km\,s^{-1}}}
\def\pc{{\rm\,pc}}
\def\ncmthree{{\rm\,cm^{-3}}}
\def\ncmtwo{{\rm\,cm^{-2}}}
\def\msun{{\,M_\odot}}
\def\gms{{\rm\,g\,s^{-1}}}
\def\emiss{{\rm\,erg\,s^{-1}\,cm^{-3}}}

\newcommand{\mdot}{\mbox{$\dot M$}{}}
\def\refindent{\par\noindent\hangindent=3pc\hangafter=1 }
\def\aa#1#2#3{\refindent#1, A\&A, {#2}, #3}

\def\aj#1#2#3{\refindent#1, AJ, {#2}, #3}
\def\apj#1#2#3{\refindent#1, {\it ApJ}, {#2}, #3.}
\def\apjlett#1#2#3{\refindent#1, {\it ApJ (Letters)}, {#2}, #3.}
\def\apjsup#1#2#3{\refindent#1, ApJS, {#2}, #3}

\def\baas#1#2#3{\refindent#1, BAAS, #2, #3}

\def\mnras#1#2#3{\refindent#1, {\it MNRAS}, {#2}, #3.}
\def\nature#1#2#3{\refindent#1, {\it Nature}, {#2}, #3.}

\def\refbook#1{\refindent#1}

\begin{document}
\setcounter{figure}{0}
\centerline{Submitted to the Editor of the Astrophysical Journal}
\bigskip
\title{Stellar Gas Flows Into A Dark Cluster Potential At The Galactic 
Center}

\author{Fulvio Melia$^{\dag}$\altaffilmark{1} and
Robert Coker$^*$\altaffilmark{2} 
\affil{$^{\dag}$Physics Department and Steward Observatory, The University 
of Arizona, Tucson, AZ 85721}
\affil{$^*$Physics Department, The University of Arizona, Tucson, AZ 85721}
}
\altaffiltext{1}{Presidential Young Investigator.}
\altaffiltext{2}{NASA, GSRP Fellow.}

\begin{abstract}
The evidence for the presence of a concentration of dark matter
at the Galactic center is now very compelling.  There is no question
that the stellar and gas kinematics within $\approx 0.01$ pc is dominated
by under-luminous matter in the form of either a massive black hole, 
a highly condensed distribution of stellar remnants, or a more
exotic source of gravity.  The unique, compact radio source Sgr A*
appears to be coincident with the center of this region, but its
size (less than about $3\times 10^{14}$ cm at $\lambda$1.35 cm) is
still significantly smaller than the current limiting volume enclosing
this mass.  Sgr A* may be the black hole, if the dark matter distribution
is point-like.  If not, we are left with a puzzle regarding its nature,
and a question of why this source should be so unique and lie only at the 
Galactic center.  In this paper, we examine an alternative to the black
hole paradigm---that the gravitating matter is a condensed cluster of
stellar remnants---and study the properties of the Galactic center wind
flowing through this region.  Some of this gas is trapped in the cluster
potential, and we study in detail whether this hot, magnetized gas
is in the proper physical state to produce Sgr A*'s spectrum.  We find
that at least for the Galactic center environment, the temperature of
the trapped gas never attains the value required for significant GHz
emission.  In addition, continuum (mostly bremsstrahlung) emission
at higher frequencies is below the current measurements and upper
limits for this source.  We conclude that the cluster potential is
too shallow for the trapped Galactic center wind to account for Sgr A*'s
spectrum, which instead appears to be produced only within an
environment that has a steep-gradient potential like that generated by
a black hole. 
\end{abstract}


\section{Introduction}
The Galactic center (GC) has long been suspected of harboring a central
mass concentration whose gravitational influence may be the cause of the
high-velocity ionized gas streamers emitting in the 12 $\mu$m [Ne II]
line (\cite{wol76}; \cite{lacy79};  for a recent review, see \cite{mez96}).
This ionized gas appears to be in orbit about a mass distribution
$\sim 3\times 10^6\;M_\odot$ within a few arcseconds 
(1$^{\prime\prime}\approx 0.04$ pc at the GC) of the compact
radio source Sgr A* (\cite{lo83}; \cite{sera85}).  A more
recent mapping of the H92$\alpha$ line emission from Sgr A West with
an angular resolution of $\sim 1^{\prime\prime}$ shows the presence of 
three dominant kinematic features, known as the Western Arc, the 
Northern Arm, and the Bar (\cite{rob93}).  The former appears to
be in circular rotation about Sgr A* at a radius of $\sim 1$ pc.  Its
velocity of $\sim 105$ km s$^{-1}$ implies that the enclosed
mass is $\sim 3.5\times 10^6\;M_\odot$.  Complementary studies of the
$\lambda2.17\;\mu$m Br$\gamma$ line emission from this region (\cite{herb93})
yield a mass of $\sim 5\;M_\odot$ for the Northern arm
and the central Bar, whose dynamics require a central concentration of $\sim
4\times10^6\;M_\odot$ within a radius of $\sim 0.17$ pc.

These early mass determinations have been supported by subsequent 
measurements of the stellar velocities and velocity dispersions
at various distances from Sgr A*.  Following
the initial work by  Sellgren et al. (1987), and Rieke \& Rieke
(1988), Haller et al. (1996) used the velocity dispersions of stars at
$\ga 0.1$ pc from Sgr A* to derive a compact mass of 
$\sim 2\times 10^6 \;M_\odot$. This is consistent with the value
of $\sim 2.5-3.2\times 10^6\;M_\odot$ derived by Genzel et al.
(1996), using the radial velocities and velocity dispersions of
$\sim 25$ early-type stars and of $\sim 200$ red giants and
supergiants within the central 2 pc.  
A third technique for tracing the central gravitational potential
is based on the acquisition of proper motions for the $\sim
50$ brightest stars within the radial range $\sim 0.004-0.4$ pc
(\cite{eck96}).  These stellar motions also seem to require a
central dark mass of $2-3\times 10^6\;M_\odot$, in good 
agreement with both the ionized gas kinematics and the velocity
dispersion measurements.

Of course, showing that the GC must contain a centralized mass
concentration does not necessarily imply that this dark matter
is in the form of a compact object with a few million solar masses.
It does not even imply that the unusual radio source Sgr A* must
be associated with it.  However, it is possible to demonstrate that
Sgr A* is probably not stellar-like.  This is based on
the fact that a heavy object in dynamical equilibrium with the 
surrounding stellar cluster will move slowly, so that a failure 
to detect proper motion in Sgr A* may be used to provide an independent 
estimate of its mass.  Using VLBI, Backer (1994) derived 
a lower mass limit of $\sim 20-2000\;M_\odot$, which appears to rule
out the possibility that Sgr A* is a pulsar, a stellar binary,
or a similarly small object (see also Reid et al. 1997).

Still, VLBA images of Sgr A* with milliarcsecond resolution (\cite{lo93}) 
show that at $\lambda$1.35 cm, its size is $2.4\pm0.2$ mas, or roughly
$2.9\times 10^{14}$ cm, much smaller than the present limiting
region within which the $2-3\times 10^6\;M_\odot$ are contained.
So the dark matter may be distributed, perhaps in the form of
white dwarfs, neutron stars, or $\sim 10\;M_\odot$ black holes
(e.g., Haller et al. 1996).  However, the latest stellar kinematic
results appear to rule out the first two possible constituents.
Genzel et al. (1996) argue that a distribution of neutron stars in 
equilibrium with the central gravitational potential should have a core
radius somewhere between $0.15$ and $0.3$ pc, significantly larger than the
value of $\le 0.07$ pc derived from the velocity data.  The same holds
true for a population of white dwarfs.  Moreover, the neutron stars
would presumably have been formed with a substantial ``kick'' and would
likely not remain bound to the nucleus (e.g., \cite{hall96}).  Thus, as long
as the dark matter distribution is in equilibrium, and we ignore the 
dynamical effects of core collapse, the only viable alternative to the
massive black hole paradigm may be a distributed population of $\sim 
10\;M_\odot$ black holes.  Although these objects will have 
formed over a Hubble time within a 
much larger volume, some cluster evolution calculations (\cite{lee95}) 
have demonstrated that due to the large mass ratio of the stellar
black holes to that of the normal stars, a core collapse could have driven
the central component of the former to very high densities. Whether
or not such a concentration is stable against mergers that would
eventually produce a single massive object is still an open question.

Whatever the composition of a distributed mass concentration is,
one would then be left with the task of accounting for the nature of
Sgr A* itself, without the benefit of invoking the deep gravitational
potential well of a point-like object (\cite{Me94}).  In this paper, 
we study in detail the emission characteristics that would be exhibited 
by a hot, magnetized plasma ``trapped'' within the dark cluster potential.
In addition to the large scale gaseous features described above, there is 
ample observational evidence in this region for the existence of 
rather strong winds in and around Sgr A* itself. The key
constituents of this morphology appear to be the cluster of mass-losing, 
blue, luminous stars comprising the IRS 16 assemblage, which is located 
within several arcseconds from Sgr A*.  Measurements of high outflow 
velocities associated with IR sources in Sgr A West (\cite{K91}) and in IRS
16 (\cite{G91}), the $H_2$ emission in the circumnuclear disk (CND) from 
molecular gas being shocked by a nuclear mass outflow (\cite{G86}), broad
Br$\alpha$, Br$\gamma$ and He I emission lines from the vicinity of IRS 16
(\cite{HKS82}; \cite{AHH90}; \cite{G91}), and radio continuum observations 
of IRS 7 (\cite{YM91}), provide clear evidence of a hypersonic wind, with a 
velocity $v_w \sim500-1000$ km s$^{-1}$, a number density $n_w\sim10^{3-4}$
cm$^{-3}$, and a total mass loss rate $\dot M_w\sim3-4\times10^{-3}\;\dot
M_\odot$, pervading the inner parsec of the Galaxy.  If the dark matter is
distributed, it is likely that a portion of this wind is captured by the
compact cluster and that it settles within its potential well.  Although
such a configuration would not include the steep gradients associated with a
black hole's gravitational field, this trapped gas (if hot and magnetized)
might conceivably still account for at least some of Sgr A*'s radiative
characteristics.  It is this possibility that we study here.

In \S\ 2 of this paper, we begin with a simple hydrostatic model of the gas 
bound to a compact dark cluster, whose mass and size are chosen to comply 
with the observed stellar kinematics.  Our subsequent 3-dimensional 
simulations of the gas flow through this mass distribution are described in
\S\ 3, and the results are presented in \S 4.  We provide an overview of our
analysis and the implications for the nature of Sgr A* in \S 5. 

\section{A Simple Hydrostatic Model Of The Gas Bound To The Dark Cluster}
Before embarking on a full 3-dimensional hydrodynamical simulation of the gas
flow through the dark cluster potential, let us briefly set up a simple 
hydrostatic model to examine the physical conditions expected for the
captured plasma.  For simplicity, we adopt a King model potential for the
dark cluster.  Later, for the hydrodynamical simulations, we will use an
$\eta$-model, described below.  The central mass density of the 
cluster is
\begin{equation}
\rho_*={9\over 4\pi u_K}{M_{cluster}\over r_K^3}\;,
\end{equation}
where $u_K$ is a number of order unity that depends on how condensed the 
stellar distribution is, and $r_K$ is the King scale radius.  If the tidal
radius $r_t$ is $2r_K$, then $u_K=0.74141$.  In this simple toy model, we
place $10^6\;M_\odot$ of stars within $1$ core radius, corresponding roughly
to $0^{\prime\prime}.02$ in projected angular size.  This constitutes a
linear core size $r_K=170$ AU's and thus $r_t=340$ AU's.  The total
integrated mass of this cluster is then $1.22\times 10^6 \;M_\odot$, which
gives $\rho_*=1.43\times 10^{-7}$ g cm$^{-3}$.

The gravitational potential per unit mass, measured relative to the center of
the cluster, is
\begin{equation}
U^{\prime}={2\pi\over 3}G\rho_*r^2\;.
\end{equation}
For the gas (with density $\rho$) to be in hydrostatic equilibrium within
this potential, its pressure must scale according to
\begin{equation}
{dP\over dr}=-\rho {dU^{\prime}\over dr}\;,
\end{equation}
where we assume an ideal equation of state (i.e., $P_g=R_g\rho T/\mu$)
and take the total pressure to be $P_g$ plus that of the equipartition
magnetic field.  Here, $R_g$ is the gas constant and $\mu$ is the mean 
molecular weight.

It is straightforward to integrate Equation (3) for a power-law temperature
distribution, 
\begin{equation}
T=T_0\left({r\over r_0}\right)^\alpha\;,
\end{equation}
which gives
\begin{equation}
P=P_0\exp\left\{-\left({r\over\bar r}\right)^{2-\alpha}\right\}\;,
\end{equation}
where
\begin{equation}
\bar r\equiv 
\left\{{2-\alpha\over 2}\,{r_s^2\over r_0^\alpha}\right\}^{1/(2-\alpha)}\;,
\end{equation}
and the pressure scale height is
\begin{equation}
r_s\equiv \sqrt{{6R_gT_0\over 2\pi\mu G\rho_*}}\;.
\end{equation}
For the gas to be confined within the core of the cluster, we must have
\begin{equation}
r_s=\eta r_K\;,
\end{equation}
where $\eta$ is of order one. This places an upper limit on the temperature,
since clearly a hotter gas will have a greater extension.  More specifically,
we see from Equation (7) that
\begin{equation}
T_0={2\pi\mu G\rho_*\eta^2\over 6R_g}\;.
\end{equation}
For example, when $\alpha=-1$, we see that 
$T_0=5\times 10^8\eta^2$ K. In 
principle, the gas can therefore attain the temperature ($\simeq 5\times 10^9$ K)
required to produce the observed GHz (thermal) synchrotron radiation from
Sgr A* if $\eta\simeq 3$ (see Melia 1994).  In addition, we see that for a
captured gas mass $M_{gas}\sim 10^{-3}\;M_\odot$, the equipartition magnetic
field can reach an intensity of $\sim 10$ gauss, or greater.  This is low
compared with the value ($\sim$ hundreds of gauss) used in the best fits by
Melia (1994) and Narayan, Yi, and Mahadevan (1995), but is certainly within the
range where an attempt to fit the spectrum with this configuration of
trapped gas is interesting. 

Thus, the physical conditions in this environment (i.e., the 
density, temperature and magnetic field) are close to the range required 
for the gas to emit cyclotron/synchrotron radiation at GHz 
wavelengths. A potential problem, however, is that the dark cluster
potential is shallow inside the core, which makes the state variables depend
only weakly on $r$, unlike the steep gradients that are apparently required by
a spectral fit with a superposition of many thermal synchrotron components.
Nonetheless, the fact that the gas is not in hydrostatic equilibrium may
introduce steeper gradients due to its dynamic structure, and so definitive
conclusions regarding the viability of this model to produce the GHz 
spectrum of Sgr A* must be based on more detailed hydrodynamic simulations,
which we describe next.

\section{The Physical Setup}

In the absence of any outflow, many of Sgr A*'s radiative characteristics 
should be due to the deposition of the energy in the Galactic center wind
into the central well.  In the classical Bondi-Hoyle (BH) scenario 
(\cite{BH44}), the mass accretion
rate for a uniform hypersonic flow past a centralized mass is
\begin{equation}\label{mdot}
\dot M_{BH} = \pi {R_A}^2 m_H n_w v_w\;,
\end{equation}
where $R_A \equiv 2 G M / {v_w}^2$ is the accretion radius and $M$ is the gravitating
mass.  At the Galactic center, for the conditions described in the Introduction,
we would therefore expect an accretion rate
$\dot M_{BH} \sim 10^{21-22} \gms$, with a capture radius $R_A \sim 0.01-0.02 \pc$.
Since this accretion rate is sub-Eddington
for a $\sim$ one million solar mass concentration, the accreting gas is mostly
unimpeded by the escaping radiation field and is thus essentially in hydrodynamic 
free-fall starting at $R_A$.  Our initial numerical simulations of this process 
(for a point object), assuming a highly simplistic uniform flow 
(\cite{RM94}; \cite{CM96}) have verified these expectations.

\subsection{The Wind Sources}

The Galactic center wind, however, is unlikely to be uniform since many stars contribute
to the mass ejection. So for these calculations, we assume that the early-type stars 
enclosed (in projection) within the Western Arc, the Northern Arm, and the Bar produce 
the observed wind.  Thus far, 25 such stars have been identified (\cite{genz96}), 
though the stellar wind characteristics of only 8 have been determined from their
He I line emission (\cite{N97};  see Table 1).  Two of those sources, IRS 13E1 and IRS 7W, seem to 
dominate the mass outflow with their high wind velocity ($\sim 1000 \kms$) and a mass 
loss rate of more than $2\times10^{-4}\;M_\odot$ yr$^{-1}$ each.
Unfortunately, the temperature of the stellar winds is not well known, and
so for simplicity we have assumed that all the winds are Mach 30; this corresponds to 
a temperature of $10^{4-5}$K.  In addition, for the sources that are used in these calculations,
their location in $z$ (i.e., along the line of sight) is determined randomly with the
condition that the overall distribution in this direction matches that in
$x$ and $y$.  With this proviso, all these early-type stars are located within the central
parsec surrounding Sgr A*.  For the calculations reported here, the
sources are assumed to be stationary over the duration of the simulation.
The stars without any observed He I line emission have been assigned
a wind velocity of $750 \kms$ and an equal mass loss rate chosen such that the total mass 
ejected by the 14 stars used here is equal to $3\times10^{-3}\;M_\odot$ 
yr$^{-1}$.  In Figure 1, we show the positions (relative to Sgr A*) of these wind sources; 
the size of the circle marking each 
position corresponds to the relative mass loss rate (on a linear scale) for that star.  
Note that although we have matched the overall mass outflow rate
to the observations, we have only used 14 of the 25 stars in the sample.
There are two principal reasons for this: (1) stars further away than
10 arcsec (in projection) from Sgr A* are outside of our volume of
solution and therefore could not be included, and (2) due to our
computational resolution limits, we needed to avoid excessively
large local stellar densities.  So small clusters of adjacent stars
were replaced with single wind sources.

\subsection{The Dark Cluster Potential}

Following Haller \& Melia (1996), we will represent the gravitational
potential of the dark cluster with an ``$\eta$-model'' (\cite{T94}).
This function represents an isotropic mass distribution with a single 
parameter, so that the mass enclosed within radius $\tilde r$ (in dimensionless
units) is given by
\begin{equation}
M_\eta(\tilde r) \propto {{{\tilde r}^\eta}\over{(1+{\tilde r})^\eta}}\;.
\end{equation}
We here restrict our examination to the case $\eta=2.5$ since this
provides the closest approximation to a King model that is physically
realizable (i.e., a nonnegative distribution function), and we scale
the mass so that $2\times10^6\;M_\odot$ are enclosed within 0.01 pc 
(see Fig. 2). With this, the total integrated mass of the dark cluster 
is $2.7\times10^6\;M_\odot$.
Thus, writing $\tilde r$ in units of $R_A$ (i.e., $r\equiv s\tilde r\times R_A$), 
and choosing $s$ to yield the observed enclosed mass at $0.01$ pc (see Fig. 2), 
we get
\begin{equation}
M_\eta(r) = 2.7\times10^6 \left({{13.84r}\over{1+13.84r}}
\right)^{5/2} \;M_\odot\;.
\end{equation}
A more recent assessment of the enclosed mass (\cite{GEOE97}) places 
a yet more rigorous constraint on the possibility of a distributed dark
matter component.  These newer observations may indeed invalidate the idea
that any realistic stable distribution of small mass objects can 
account for the observed potential.  In the future, it may be of
interest to carry out a calculation similar to that reported here,
but for a more condensed dark matter distribution consistent with
this latest observation.  As we shall see, however, the present hydrodynamical 
calculation strongly suggests that a cluster potential
cannot easily reproduce Sgr A*'s spectrum, and this next generation
of calculations may therefore be unnecessary. 

In a complex flow, generated by many wind sources, the wind
velocity and density are not uniform, so the accretion radius may not
be independent of angle.  To set the length scale for the
simulations, we shall therefore adopt the value $R_A = .018 \pc$
(for which 1$^{\prime\prime}$ = 2.3 $R_A$) as a reasonable
mean representation of this quantity.  

\subsection{The Hydrodynamical Modeling}

\subsubsection{The Hydrodynamics Code}

We use a modified version of the numerical algorithm ZEUS, a general 
purpose code for MHD fluids developed at NCSA (\cite{SN92}; \cite{No94}).  
The code uses Eulerian finite differencing with the following relevant 
characteristics: fully explicit in time; operator and directional 
splitting of the hydrodynamical variables; fully staggered grid; 
second-order (van Leer) upwinded, monotonic interpolation for advection; 
consistent advection to evolve internal energy and momenta; and explicit 
solution of internal energy.  More details can be found in the references.
The code was run on the massively parallel Cray T3E at NASA's Goddard 
Space Flight Center under the High Performance Computing Challenge program.  
The production run for the calculation presented here used 125 T3E 
processors and ran at more than 8 Gflops.  The code assigns each processor 
to a ``tile'' and each tile consists of $60^3$ zones.  Zone sizes are 
geometrically scaled by a factor of 1.02 so that the central zones are 
$\sim 20$ times smaller than the outermost zones, mimicking
the ``multiply nested grids'' arrangement used by other researchers 
(e.g., \cite{RM94}).  This allows for maximal resolution of the central 
region (within the computer memory limits available) while sufficiently 
resolving the wind sources and minimizing zone-to-zone boundary effects. 
The total volume is $(40 R_A)^3$ or $\sim(0.7 \pc)^3$ with the center of
the dark cluster distribution being located at the origin.

The density of the gas initially filling the volume of solution is set to 
a small value and the velocity is set to zero.  In order to reach 
equilibrium more quickly, the internal energy density is chosen such that 
the initial temperature is $\sim10^2$ K.  Free outflow conditions are 
imposed on the outermost zones and each time step is determined by the 
Courant condition with a Courant number of 0.5.  The 14 stellar wind 
sources are modeled by forcing the velocity in 14 subregions of 125 zones 
each to be constant with time while the densities in these subvolumes are 
set so that the total mass flow into the volume of solution, $\dot M_w$, 
is given by Table \ref{srcs}.  Also, the magnetic field of the winds
is assumed to always be at equipartition.  The angular momentum and mass accretion 
rates reported in the next section are calculated by summing the relevant 
quantity in zones located within $0.1 R_A$ of the origin.

\subsubsection{Heating and Cooling}

Although we are modeling the wind sources more realistically than in previous 
work, we here ignore the effects of the magnetic field on the large scale kinematics.
We take the medium to be an adiabatic polytropic gas, with $\gamma = 5/3$.  Building on
previous work (\cite{Me94} and \cite{CM97}), we have included a first order 
approximation to magnetic dissipative heating as well as an accurate expression
for the cooling due to magnetic bremsstrahlung, thermal bremsstrahlung, line emission,
radiative recombination, and 2 photon continuum emission for a gas with cosmic
abundance. For magnetic heating, we assume that the magnetic field never rises 
above equipartition.  If compression and flux conservation would otherwise
dictate a magnetic field larger than the equipartition value, the field lines 
are assumed to reconnect rapidly, converting the magnetic field energy into thermal
energy, thereby re-establishing equipartition conditions.  In the future, we will 
use a more detailed treatment of this dissipation process based on the scheme
described in Kowalenko \& Melia (1997).  

The cooling function includes a multiple-Gaussian fit to the relevant
cooling emissivities provided by N. Gehrels (see \cite{GW93} and references cited 
therein), though with the thermal bremsstrahlung portion supplanted
with more accurate expressions that are valid over a broader range of
physical conditions and with the inclusion of magnetic bremsstrahlung.  
For the former, we use (\cite{RL79}) 
\begin{equation}
\epsilon_{TB} = 2.84\times10^{-27} n_e^2 T^{1/2} \emiss\;,
\end{equation}
with a Gaunt factor of 1.2 and a Z of 1.3 (\cite{RL79}),
while for the latter (\cite{Me94}) 
\begin{equation}
\epsilon_{MB} = 1.06\times10^{-15} G(T) n_e B^2 \emiss\;,
\end{equation}
where $n_e$ is the electron number density, $T$ is the temperature,
$B$ is the magnetic field, and $G(T)$ is an analytical approximation
to Equation (27) in Melia (1994) (good to better than 5\%), given by
the expression
\begin{equation}
G(T) = {{3X^2+12X+12}\over{X^3+X^2}}\;,
\end{equation}
where $X = m_e c^2/kT$.
The ionization fraction, $\Delta$, is found by balancing recombination with collisional
ionization (\cite{Ros97}) and setting $n_e = \Delta n_H$, where $n_H$ is the Hydrogen
number density.  This involves solving Equation (1e) in Rossi et al. (1997) for 
$\Delta$ assuming that the left hand side is zero:  
\begin{equation}
\Delta = {{1}\over{1+0.445 T^{-1}e^{157890/T}}}\;.
\end{equation}
In addition, when calculating $n_e$, we ensure that radiative cooling ceases 
below $3\times10^3$K; it is likely that the assumption of steady state equilibrium, 
made in deriving the cooling rates in Gehrels \& Williams (1993), is invalid below 
this temperature.  Figure 3 shows the resulting emissivities as functions of $T$ 
for a magnetic field of 10 milliGauss and a Hydrogen number density 
$n_H = 10^4 \ncmthree$.  Note that cooling due to Comptonization and any pair production 
have not yet been included since they are not thought to be significant in the vicinity 
of Sgr A*.  Also, it is assumed that the optical depth is small throughout the volume
of solution.

\subsection{Calculation of the Spectrum}

In order to calculate the observed continuum spectrum, we assume that the observer is
positioned along the negative $z$-axis at infinity and we sum the emission from all zones that 
are located at a projected distance, $R_{xy}$, of less than $0.1\,R_A$.  Further, we assume
that scattering is negligible and that the optical depth is less than unity.  At the
temperature and density that we encounter here, the dominant components of the continuum 
emissivity are electron-ion ($\epsilon_{ei}$) and electron-electron ($\epsilon_{ee}$) 
bremsstrahlung.  Due to the interstellar medium's weak magnetic field ($\sim\mu G$), 
the emissivity from magnetic bremsstrahlung, even at small $r$, is orders of magnitude 
smaller than $\epsilon_{ei}$ and $\epsilon_{ee}$.  
In the non-relativistic limit, we use $\epsilon_{ei}^{NR}$ and $\epsilon_{ee}^{NR}$
based on the expressions in Gould (1980) and Gould (1981), though excluding the Sommerfeld
and other higher order corrections.  Defining $E=h\nu/m_e c^2$ and $\beta=h\nu/2k_B T$,
we have 
\begin{equation}
\epsilon_{ei}^{NR} = C_{ei}^{NR} ( 1 + Ef/4) e^{-\beta} k_0(\beta) \sqrt{X} n_e^2\;,
\end{equation}
where $f = 1/4\beta+k_1(\beta)/k_0(\beta)$, $k_0$ and $k_1$ are the 0th
and 1st order modified Bessel functions, respectively, and $C_{ei}^{NR}$ is a constant
given by
\begin{equation}
C_{ei}^{NR} = {{4}\over{3\pi}}\sqrt{{{2}\over{\pi}}} c h \alpha^3 (\lambda/2\pi)^2,
\end{equation}
where $\alpha$ is the fine-structure constant, $\lambda$ is the electron's Compton 
wavelength, $c$ is the speed of light, and $h$ is Planck's constant.  Here, and
in the following expressions, the units of $\epsilon$ are erg s$^{-1}$ Hz$^{-1}$ 
steradian$^{-1}$ cm$^{-3}$ Similarly, we have
\begin{equation}
\epsilon_{ee}^{NR} = C_{ee}^{NR} \sqrt{T} \beta e^{-\beta} k_0(\beta) g n_e^2\;,
\end{equation}
where
\begin{equation}
g = {{3}\over{4\beta}} + {{k_1(\beta)}\over{k_0(\beta)}} + {{0.9 e^{-\beta} k_0(\beta)}\over{\beta}}\;,
\end{equation}
and
\begin{equation}
C_{ee}^{NR} = {{128}\over{15}} \alpha^3 \lambda^2 h \sqrt{{{k_B}\over{\pi m_e}}}\;.
\end{equation}
For the relativistic electron-ion bremsstrahlung emissivity, we use an expression
from Quigg (1968).  Defining $\beta_1\equiv h\nu/k_B T$, then we have
\begin{equation}
\epsilon_{ei}^{ER} = C^{ER} {{k_2(X)}\over{X^2}} e^{-\beta_1}
\left( {{40}\over{3}} + {{4\beta_1}\over{3}} -\beta_1^2 + \left( {{16}\over{3}} + {{8\beta_1}\over{3}} + 
2\beta_1^2 \right)
  \left(\log{{{2}\over{X}}}-\gamma\right) +
  e^{\beta_1} Ei(-\beta_1) \left( -{{16}\over{3}} + {{8\beta_1}\over{3}} - 2 \beta_1^2 \right)
  \right) n_e^2
\end{equation}
where $Ei$ is the exponential integral, $k_2$ is the 2nd order modified Bessel function,
$\gamma$ is Euler's constant and $C^{ER}$ is a constant given by
\begin{equation}
C^{ER} = {{r_0^2 \alpha c h}\over{2 \pi}}\;,
\end{equation}
with $r_0$ the classical electron radius.
For the relativistic electron-electron bremsstrahlung emissivity, we use an expression
from Alexanian (1968):
\begin{equation}
\epsilon_{ee}^{ER} = C^{ER} e^{-\beta_1} \left( {{28}\over{3}} + 2\beta_1 + {{\beta_1^2}\over{2}} +
  \left( {{16}\over{3}} + {{8\beta_1}\over{3}} + 2\beta_1^2 \right) \left( \log{{{2}\over{X}}}-\gamma\right)
     -e^{\beta_1} Ei(-\beta_1) \left( {{8}\over{3}} - {{4\beta_1}\over{3}} +\beta_1^2 \right) \right) n_e^2.
\end{equation}
In the transrelativistic region ($X\sim1$), we use a weighted average of the NR and ER expressions.
Note that the above equations are slightly different from those in the references due to a number of
typographical errors in the originals. 
Taking the effects of refraction into account, the final calculated luminosity is
\begin{equation}
L_{\nu} = 4\pi \sum_{ij}^{R_{ij}<1 R_A} \sum_k (\epsilon_{ei}+\epsilon_{ee})
\sqrt{1 - {{e^2 n_{e,ijk}}\over{\pi m_e\nu^2}}}\;dV_{ijk}\; e^{-\sum_{1}^{k} d\tau_k}\;,
\end{equation} 
where $k$ is the zone number along the line of sight, $d\tau_k$ is its optical
depth and $n_{e,ijk}$ is the electron number density within the zone $(ijk)$.  
Refraction is important only below $\sim 0.2$ GHz, where it accounts for
the low-frequency turnover evident in the spectra shown in Figure 8.

\section{Results}

The large scale gas morphology through the central
dark cluster toward the end of the hydrodynamical simulation 
is shown in Figures 4 and 5.  Figure 4 is a plot of the column density
viewed along the $z$-axis while Figure 5 shows the emitted intensity (i.e.,
the integrated emissivity along the line-of-sight).  Of particular
interest in these images is the appearance of streaks of high-velocity
gas (``streamers'') that are very reminiscent of features, such as the
so-called ``Bullet'' seen near the Galactic center (Yusef-Zadeh, et al. 1996).
In our simulation, these structures are produced predominantly within the
wind-wind collision regions, and in the future, we shall consider in 
greater detail the possibility that the observed high-velocity gas
components near Sgr A* are produced in this fashion. Note in these figures
the relative locations of the wind-producing stars (see also Fig. 1), and
the dominant role played by IRS 13E1 to the lower right of the compact
radio source.

It should also be noted that the gas distribution in a multiple-wind source
environment like that modeled here is distinctly different from that of
a uniform flow past a central accretor (Coker \& Melia 1997).  Unlike the
latter, the former does not produce a large-scale bow shock, and therefore
the environmental impact of the gravitational focusing by the central
dark mass has significantly less order in this case.  For example, it is
less likely in a multiple-wind source environment that the interaction
between a centralized mass and the Galactic center wind can craft the
type of channeled flow required to produce the mini-cavity.  We shall
defer a more extensive discussion of this point to a later publication
in which we report the results of a multiple-wind source simulation for
the case in which the dark matter is in the form of a massive black hole
rather than a distributed dark cluster.

In Figure 6 we show the mass enclosed within a sphere of radius 0.1 $R_A$
as a function of time, centered on the midpoint of the dark cluster.
In similar fashion, Figure 7 is a plot of the enclosed energy within
this same volume.  After reaching equilibrium several sound crossing
times after the start of the simulation, the enclosed mass and energy begin 
to fluctuate aperiodically, reflecting the turbulent cell nature of
the flow in and out of the central region.  Typically $2.7\times 10^{-3}
\;M_\odot$ of gas (the dashed line in this figure) is trapped within the 
cluster at any given time.  Note also that although the gas is 
highly supersonic, with most of the energy in kinetic form, 
the thermal energy can be boosted rather suddenly when the enclosed
magnetic field energy is dissipated.  This occurs when strong shocks
pass through this region;  the shocks compress the field sufficiently
to the point where it reaches, or even surpasses, equipartition and
dissipation ensues. The turbulent, time-dependent nature of the flow
is clearly evident in this Figure too.

VLBA imaging of Sgr A* (Lo, et al. 1993) restricts its size at
$\lambda$1.35 cm to $2.4 \pm 0.2$ mas, corresponding to a linear
dimension of $\approx 2.9\times 10^{14}$ cm.  The smallest cell size
in our simulation is $7\times10^{14}$ cm.  To minimize the inaccuracy
due to numerical fluctuations, we have calculated the spectrum
from a central region roughly $10$ times this size ($0.1 R_A$), to include
at least $100$ zones.  Thus clearly our predicted spectrum constitutes 
an upper limit to the actual emission expected from Sgr A*.  
We have also introduced the additional simplification of ignoring
the line emission.  Since the average gas temperature within these zones is 
$\sim10^7$ K, the contribution to the luminosity due to fb and bb processes 
(see Fig. 3) is at most comparable to (though mostly less than) that due to
continuum processes.  Further, the line emission will not contribute significantly
to the radio portion of the spectrum.  For simplicity, we have
therefore only calculated the continuum spectrum from this central
region for comparison with that observed from Sgr A*.

We show the predicted spectrum for this simulation in Figure 8.  
Not surprisingly, the spectrum has characteristics very reminiscent of
a hot gas trapped in a shallow gravitational potential (e.g., Limaneto,
et al. 1997).  That is, it is an accumulation of bremsstrahlung components
with a high-energy shoulder characteristic of the highest temperature $T_{max}$
attained by the gas. In our simulation, $T_{max}$ never reached the values
required for cyclotron/synchrotron emission to become important (see \S\ 2
above). As such, the radio luminosity is more than 4 orders of magnitude 
below that actually observed from Sgr A*.  This is of course consistent with
the brightness temperature limit associated with this source.  We note also
that the flux predicted by this process at X-ray and $\gamma$-ray energies is 
significantly below that required to account for the ROSAT datum 
(Predehl \& Truemper 1994) and is well below the SIGMA upper limit
(Goldwurm, et al. 1994).

\section{Discussion}
 
Based on the simulation reported here, it does not appear that the gravitational
potential of a cluster of stellar remnants, even an extremely compact one, can 
compress the gas from stellar winds to the point where the temperature, density
and magnetic field are sufficient to drive an observationally significant
cyclotron/synchrotron emissivity at GHz frequencies.  As we speculated
earlier, this is due entirely to the fact that the cluster potential is
flat in its middle, unlike the steep gradients encountered by the infalling
gas near a black hole.  Although the spatial resolution near the origin
can be improved over that used here, it is unlikely that this improvement
in the model can alter this deficiency of the flattened cluster potential.

Thus, aside from the issue of (i) whether an equilibrium dark cluster can
even account for the observed Galactic center potential, and (ii) whether
such a compact cluster can be stable over a significant fraction of the
age of the Galaxy, our 3D hydrodynamical simulation has demonstrated that
the radio emissivity from the gas trapped within this cluster cannot
reproduce Sgr A*'s spectrum.  The alternative viable options would therefore
appear to be (i) that Sgr A* is unrelated to the accreting or trapped
Galactic center gas, which then raises the question of why such a unique
source should lie only at the Galactic center, and (ii) that Sgr A* is
the signature of an accreting, massive black hole.

Our simulation has also raised other very interesting and important
questions that are best pursued elsewhere. These include the nature of
the high-velocity gas streamers observed in radio continuum images of 
the Galactic center, and the specifics of the environmental impact
of multiple wind sources interacting with the central black hole. 

\par
{\bf Acknowledgments} We gratefully acknowledge helpful discussions 
with Joseph Haller regarding the central cluster potential.  This work 
was partially supported by NASA grants NAGW-2518 and NGT-51637.

{}

\begin{deluxetable}{lccccc}
\tablecaption{Parameters for Galactic Center Wind Sources\label{srcs}}
\tablehead{
  \colhead{Star}
& \colhead{x\tablenotemark{a} (arcsec)}
& \colhead{y\tablenotemark{a} (arcsec)}
& \colhead{z\tablenotemark{a} (arcsec)}
& \colhead{v ($\kms$)}
& \colhead{\mdot (${10}^{-5}\;M_\odot$ yr$^{-1}$)}
}
\startdata

 IRS 16NE  & -2.6 &  0.8 &  5.5 &  550 &  9.5 \nl
 IRS 16NW  &  0.2 &  1.0 &  7.3 &  750 &  5.3 \nl
 IRS 16C   & -1.0 &  0.2 & -7.1 &  650 & 10.5 \nl
 IRS 16SW  & -0.6 & -1.3 &  4.9 &  650 & 15.5 \nl
 IRS 13E1  &  3.4 & -1.7 & -1.5 & 1000 & 79.1 \nl
 IRS 7W    &  4.1 &  4.8 & -5.1 & 1000 & 20.7 \nl
 AF        &  7.3 & -6.7 &  8.5 &  700 &  8.7 \nl
 IRS 15SW\tablenotemark{b}  &  1.5 & 10.1 &      &  700 & 16.5 \nl
 IRS 15NE\tablenotemark{b}  & -1.6 & 11.4 &      &  750 & 18.0 \nl
 IRS 29N\tablenotemark{c}   &  1.6 &  1.4 &  3.5 &  750 & 12.9 \nl
 IRS 33E\tablenotemark{c}   &  0.0 & -3.0 &  1.5 &  750 & 12.9 \nl
 IRS 34W\tablenotemark{c}   &  3.9 &  1.6 & -6.4 &  750 & 12.9 \nl
 IRS 1W\tablenotemark{c}    & -5.3 &  0.3 &  7.8 &  750 & 12.9 \nl
 IRS 9NW\tablenotemark{cd}  & -2.5 & -6.2 & -3.8 &  750 & 12.9 \nl
 IRS 6W\tablenotemark{c}    &  8.1 &  1.6 &  3.6 &  750 & 12.9 \nl
 AF NW\tablenotemark{cd}    &  8.3 & -3.1 & -2.1 &  750 & 12.9 \nl
 BLUM\tablenotemark{b}      &  9.2 & -5.0 &      &      &      \nl
 IRS 9S\tablenotemark{b}    & -5.5 & -9.2 &      &      &      \nl
 Unnamed 1\tablenotemark{b} &  1.3 & -0.6 &      &      &      \nl
 IRS 16SE\tablenotemark{b}  & -1.4 & -1.4 &      &      &      \nl
 IRS 29NE\tablenotemark{b}  &  1.1 &  1.8 &      &      &      \nl
 IRS 7SE\tablenotemark{b}   & -2.7 &  3.0 &      &      &      \nl
 Unnamed 2\tablenotemark{b} &  3.8 & -4.2 &      &      &      \nl
 IRS 7E\tablenotemark{b}    & -4.2 &  4.9 &      &      &      \nl
 AF NWW\tablenotemark{b}    & 10.2 & -2.7 &      &      &      \nl

\enddata
\tablenotetext{a}{Relative to Sgr A* in l-b coordinates where negative x
is east and negative y is south of Sgr A*}
\tablenotetext{b}{Star not used in these calculations}
\tablenotetext{c}{Wind velocity and mass loss rate fixed (see text)}
\tablenotetext{d}{Star position changed slightly due to finite physical resolution}
\end{deluxetable}
\newpage

\begin{figure}\label{fig-winds}
\epsscale{1.00}
\plotone{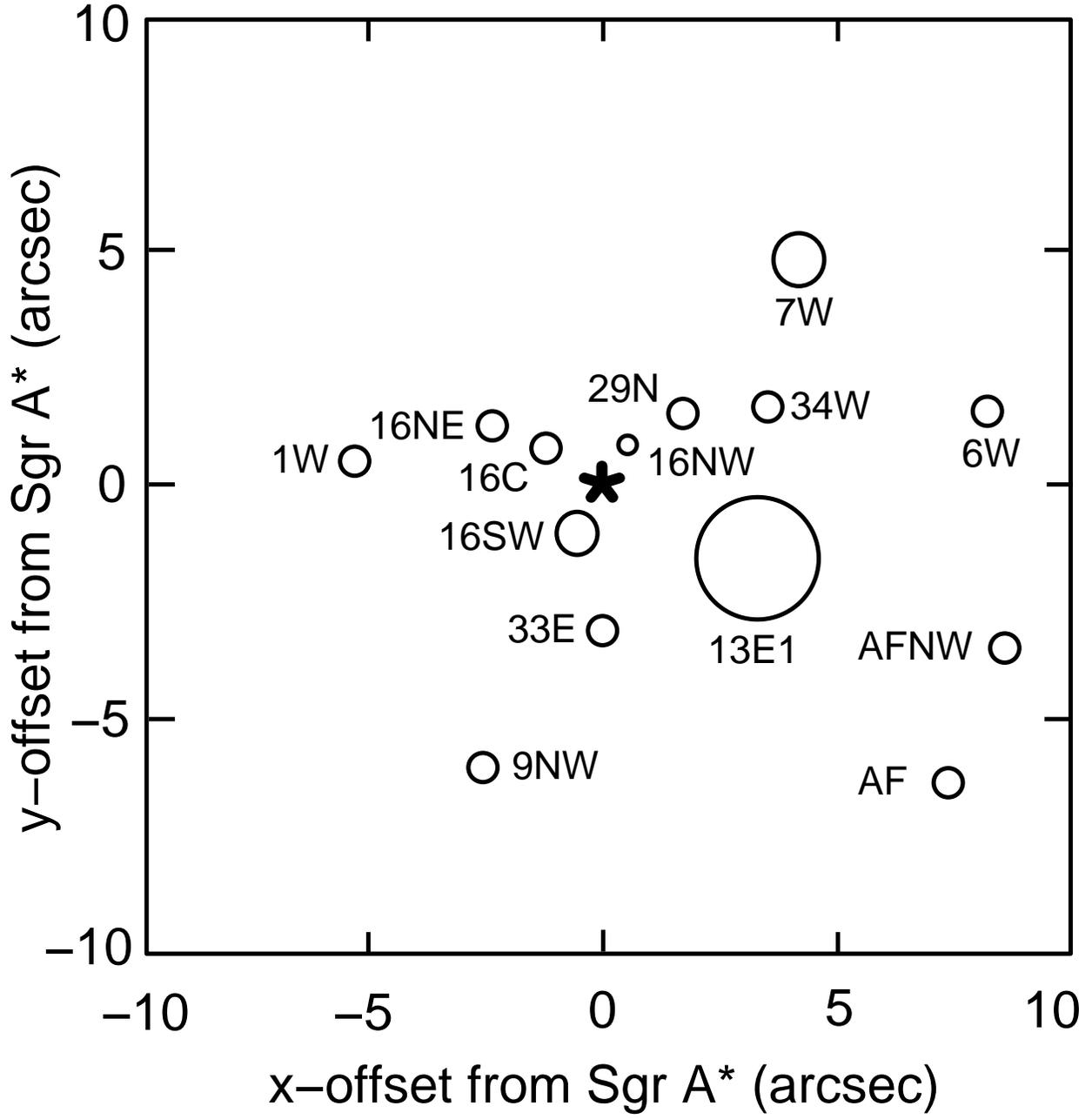}
\vskip 1.0in
\caption{Location of the 14 wind-producing stars used in the
simulation reported below, relative to the position of Sgr A*
indicated by the * symbol. The radius of each circle corresponds
(on a linear scale) to that star's mass loss rate.  Setting
the scale is 13E1, with $\dot M=7.9\times 10^{-4}\;M_\odot$ yr$^{-1}$.}
\end{figure}
\newpage

\begin{figure}\label{fig-eta}
\epsscale{1.00}
\plotone{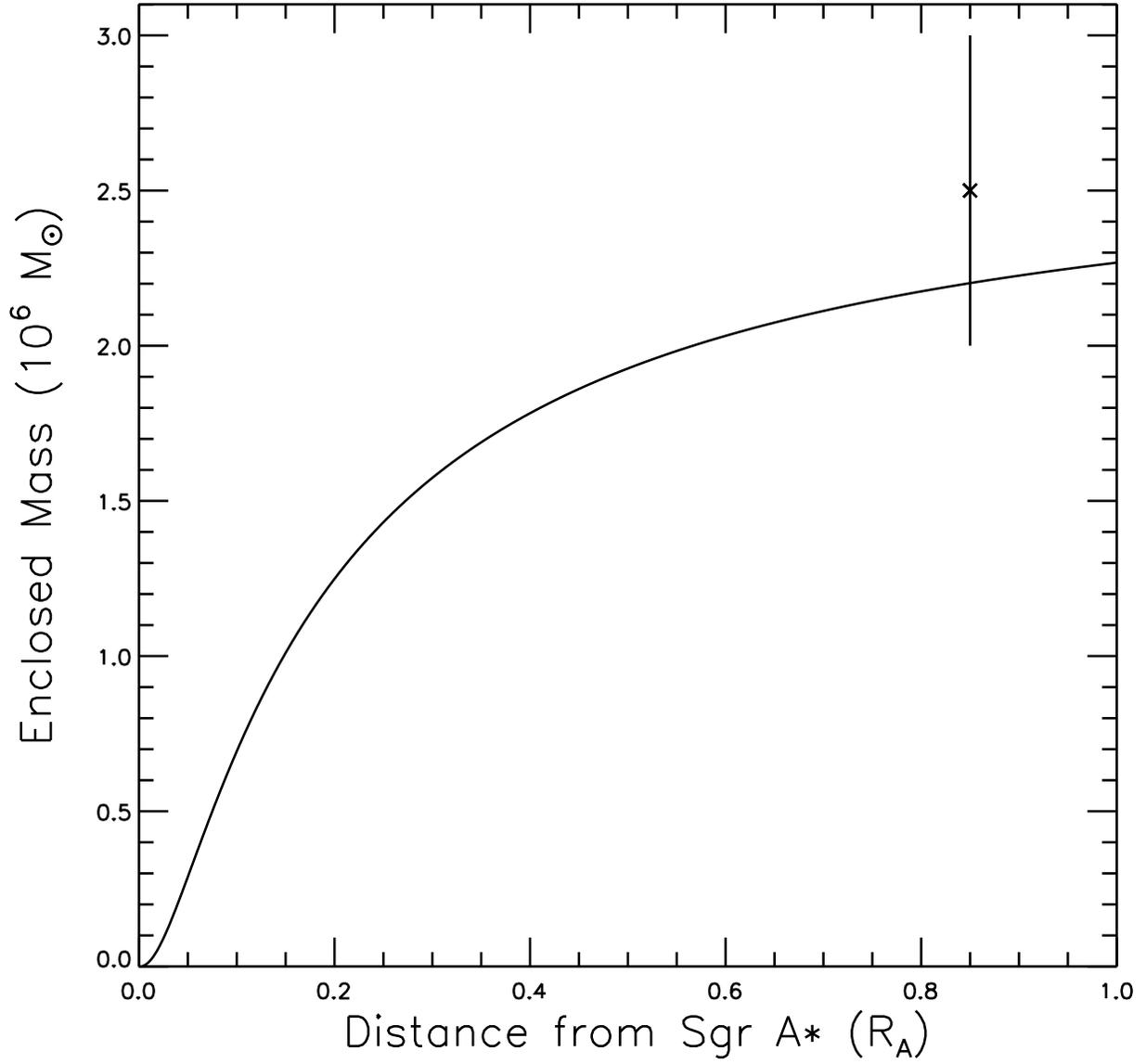}
\caption{Plot of the enclosed mass versus distance from Sgr A* in
an $\eta$-model with $\eta=2.5$.  Also shown is a recent observational
determination of the enclosed mass ($2.5 \pm 0.5 \times 10^6 \msun$)
within 0.015 pc (Eckart \& Genzel 1997).}
\end{figure}
\newpage

\begin{figure}\label{fig-emis}
\epsscale{1.00}
\plotone{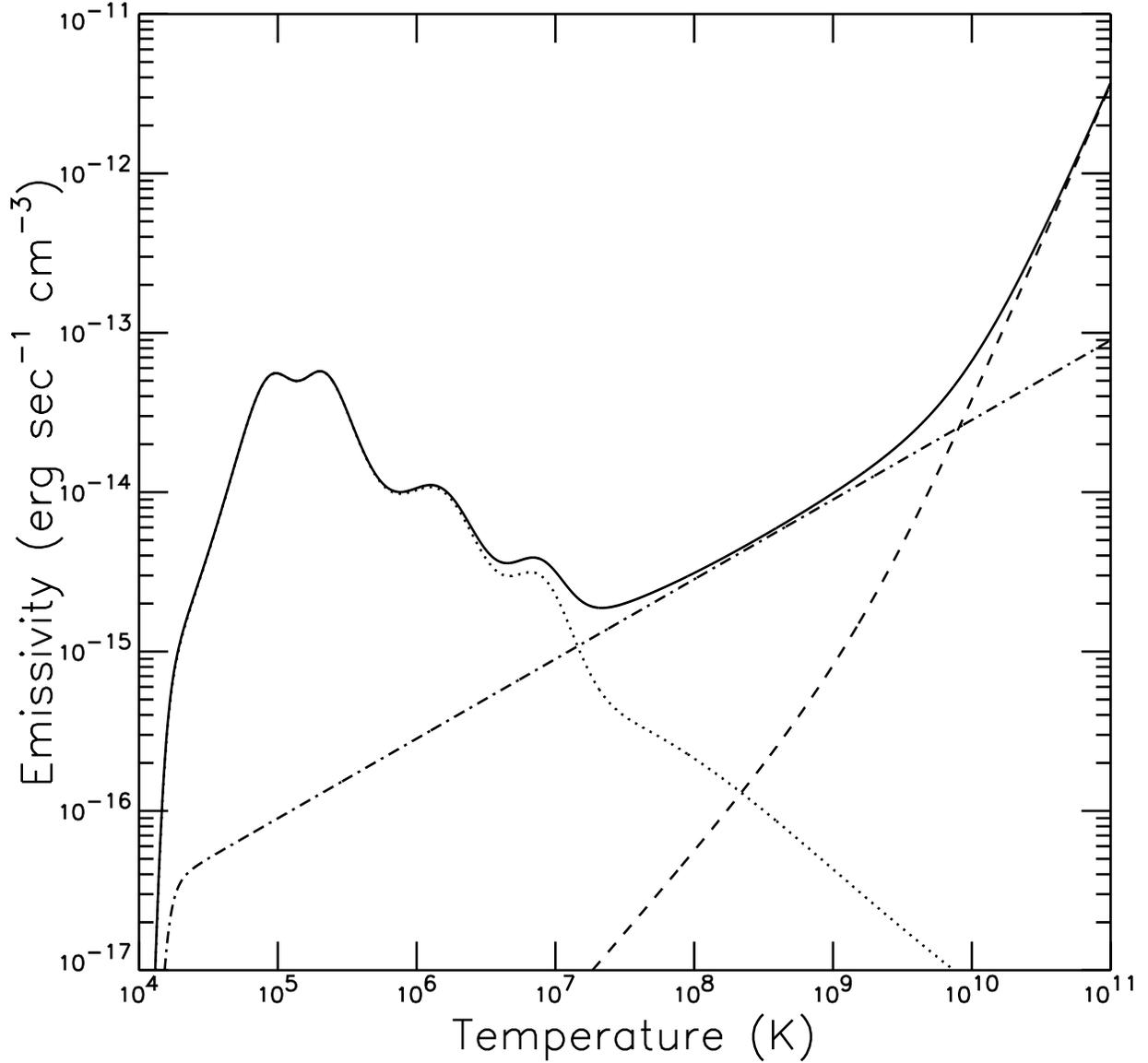}
\caption{A plot of emissivity versus temperature used in the simulations,
for the representative parameter values $B=10^{-2}$ Gauss and $n_H=10^4$
cm$^{-3}$.  The solid curve is the total, the dot-dashed curve corresponds 
to thermal bremsstrahlung, the dashed curve is for magnetic bremsstrahlung, 
and the dotted curve corresponds to the sum of the other cooling mechanisms 
(see text).}
\end{figure}
\newpage

\begin{figure}\label{fig-colden}
\epsscale{1.00}
\plotone{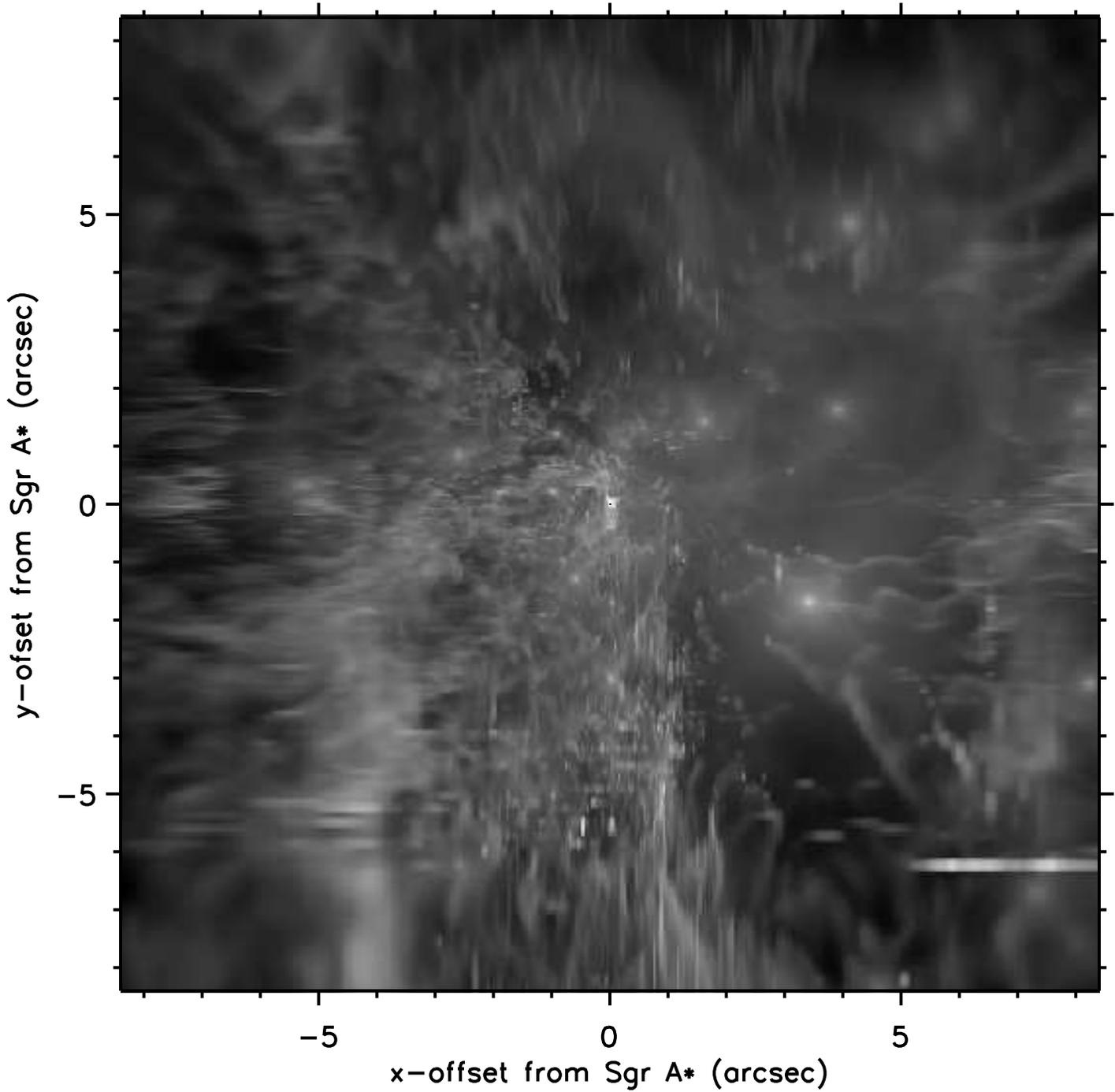}
\caption{A plot of the line-of-sight column density along the $z$-axis for the
hydrodynamical simulation at the end of the calculation (at $\sim$ 1450 years).
The grey scale is logarithmic
with solid white corresponding to a column density of $\sim1.2\times10^{21}\ncmtwo$ and
black corresponding to $\sim6.5\times10^{16}\ncmtwo$.  Sgr A* is located at the
center of the image.}
\end{figure}
\newpage
\begin{figure}\label{fig-emiss}
\epsscale{1.00}
\plotone{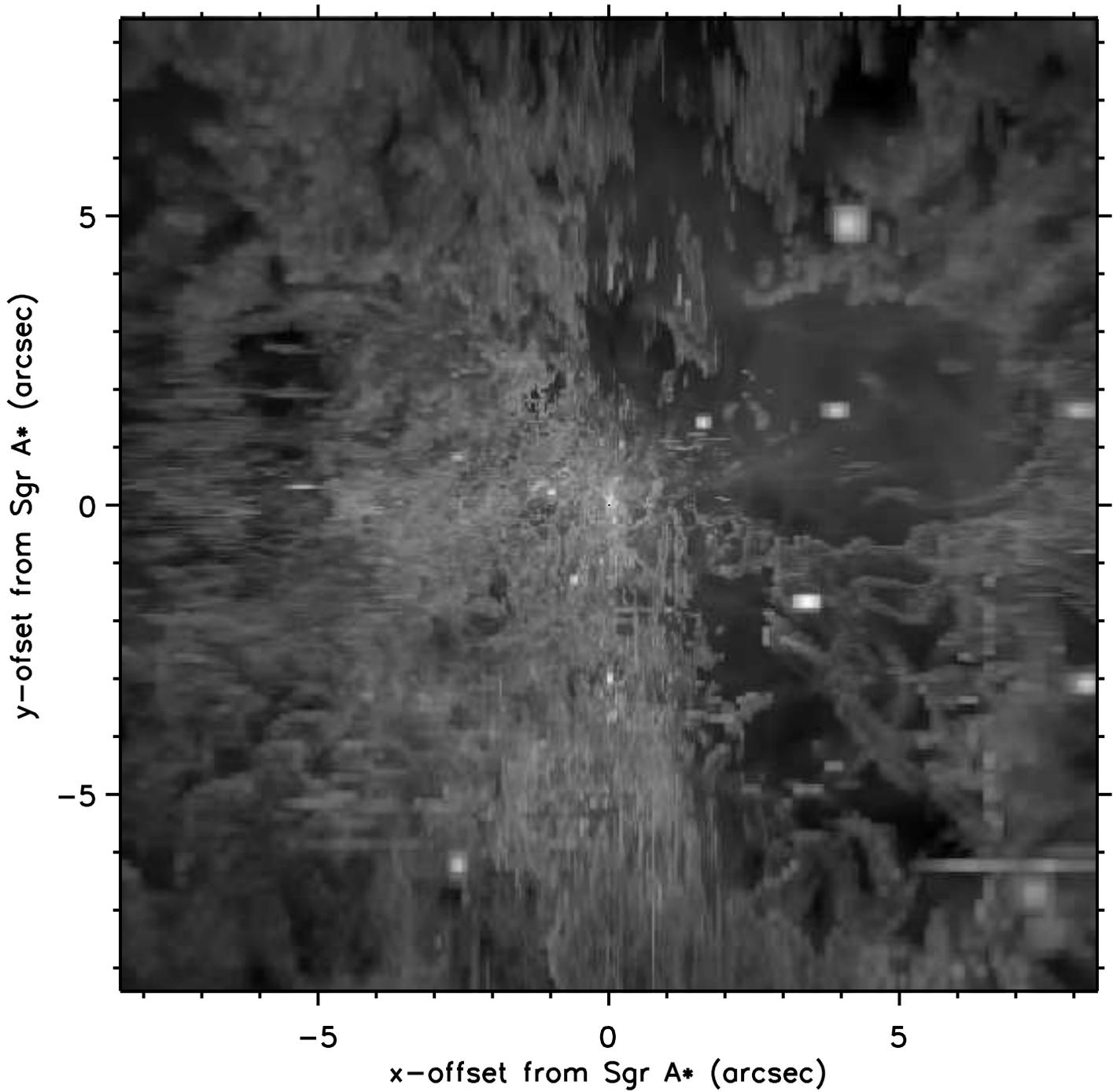}
\caption{A plot of the line-of-sight integrated emissivity along the $z$-axis for the
hydrodynamical simulation at ($\sim$ 1450 years).  For this figure, it is assumed
that there is no scattering or absorption.  The grey scale is logarithmic
with solid white corresponding to a frequency-integrated intensity of 
$\sim1.1\times10^{5}$ erg cm$^{-2}$ s$^{-1}$ steradian$^{-1}$,
and black corresponding to $\sim1\times10^{-2}$ erg cm$^{-2}$ s$^{-1}$ steradian$^{-1}$.
Sgr A* is located at the center of the image.  For reference, note that the
volume of this central region is $7\times10^{47}$ cm.}
\end{figure}
\newpage

\begin{figure}\label{fig-mass}
\epsscale{1.00}
\plotone{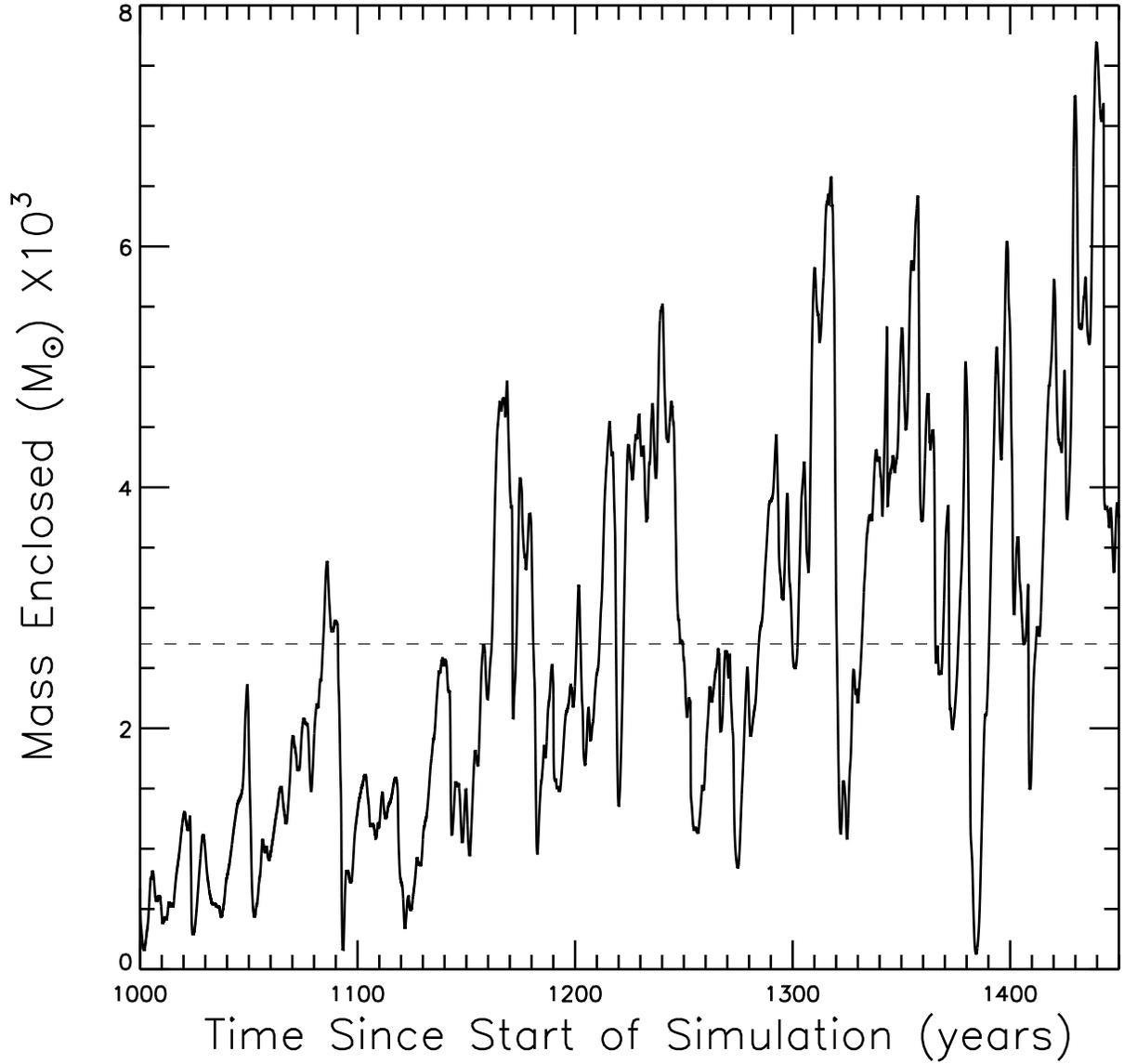}
\caption{A plot of mass enclosed within 0.1 $R_A$ versus time for the hydrodynamical simulation.
The dashed line is the average ($2.7\times10^{-3} \msun$) value for times beyond 1000 years.
Note the frequent large amplitude fluctuations over a time scale of less than a few decades.}
\end{figure}
\newpage

\begin{figure}\label{fig-ener}
\epsscale{1.00}
\plotone{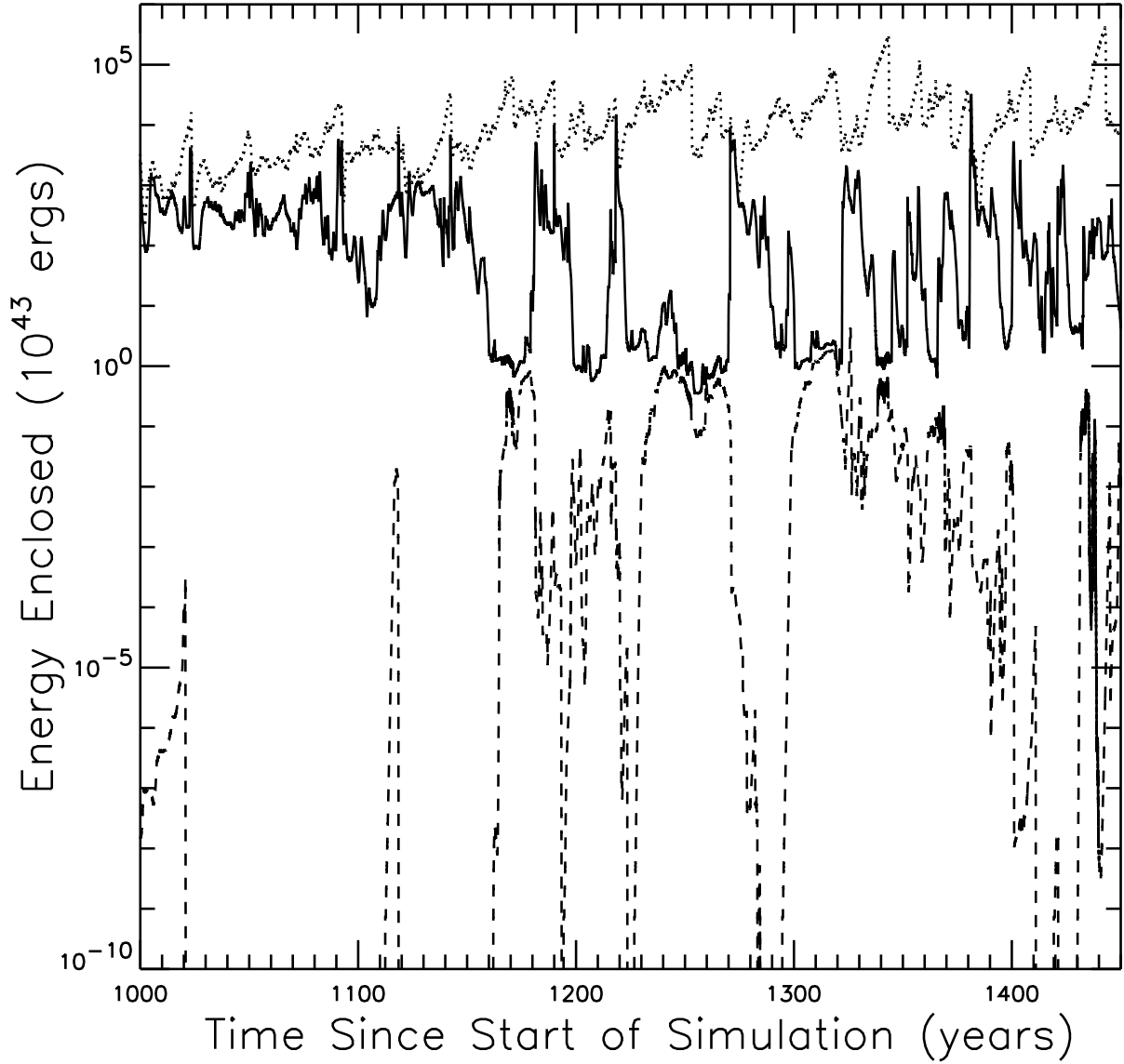}
\caption{Plot of energy enclosed within 0.1 $R_A$ versus time for the hydrodynamical simulation.
The solid, dashed, and dotted curves correspond to thermal, kinetic, and magnetic energy, respectively.
Note that the thermal energy rises whenever the magnetic energy is dissipated.  Also, peaks in
the thermal energy are often associated with rapid changes in the kinetic energy; this occurs when
shock fronts travel through the central region.}
\end{figure}
\newpage

\begin{figure}\label{fig-spec}
\epsscale{1.00}
\plotone{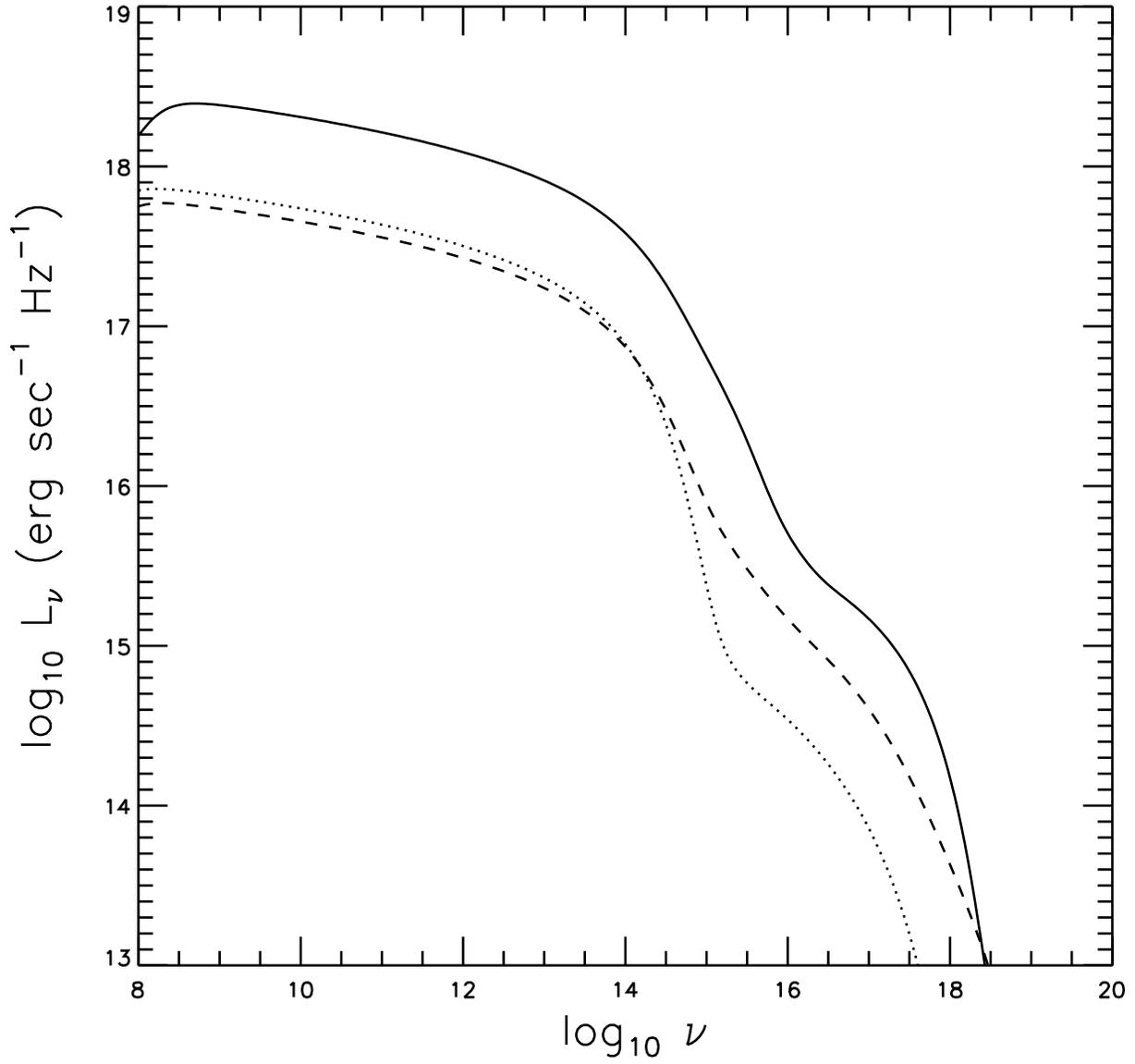}
\caption{Plot of the predicted luminosity density versus frequency for 3 points in
time during the 3D hydrodynamical simulation.  The solid, dotted, and dashed curves
are at approximately 1200, 1300, and 1400 years, respectively, after the start of the 
simulation.  Equilibrium occurs after several sound crossing times, corresponding
roughly to 1000 years.}
\end{figure}

\end{document}